\documentclass[12pt,a4paper]{article}

\usepackage[utf8]{inputenc}
\usepackage{authblk}
\usepackage{amssymb}
\usepackage{amsmath}
\usepackage{amsthm}
\usepackage{graphicx}
\usepackage{tikz}
\usepackage{setspace}
\usetikzlibrary{positioning}
\def\gev{{\hbox{GeV}}}
\def\mev{{\hbox{MeV}}}

\def\alr{A_{\rm LR}}
\def\afb{A_{\rm FB}}

\def\ptop{P_{t}}

\def\sba{$\sigma^{\text{Born}}_{\alpha(0)}$}
\def\sbg{$\sigma^{\text{Born}}_{G_\mu}$}

\def\dbg{$\delta^{\text{Born}}_{G_\mu / \alpha(0)}$}

\def\swa{$\sigma^{\text{weak}}_{\alpha(0)}$}
\def\swg{$\sigma^{\text{weak}}_{G_\mu}$}

\def\swha{$\sigma^{\text{weak+ho}}_{\alpha(0)}$}
\def\swhg{$\sigma^{\text{weak+ho}}_{G_\mu}$}

\def\dwg{$\delta^{\text{weak}}_{G_\mu / \alpha(0)}$}

\def\dwhg{$\delta^{\text{weak+ho}}_{G_\mu / \alpha(0)}$}

\title{Electroweak radiative corrections to 
polarized top quark pair production}
\author[1]{A.\,Arbuzov}
\author[1]{S.\,Bondarenko}
\author[2]{L.\,Kalinovskaya}
\author[2]{R.\,Sadykov}
\author[2,3]{V.\,Yermolchyk}

\affil[1]{\small Bogoliubov Laboratory of Theoretical Physics, JINR, 141980, Dubna, Moscow region, Russia}
\affil[2]{\small Dzhelepov Laboratory of Nuclear Problems, JINR, 141980 Dubna, Moscow region, Russia}
\affil[3]{\small Institute for Nuclear Problems, Belarusian State University, Minsk, 220006  Belarus}
\date{{\vspace{-5ex}}}
\begin{document}
\maketitle
\begin{abstract}

Electroweak effects in the
 $e^+e^-  \to t \bar{t}$ annihilation process are described with taking into account polarization
of the initial and final particles.
We investigate the effects of complete one-loop electroweak radiative corrections (RCs) and higher-order radiative effects to the 
total cross section and
analyze different types of asymmetries for polarized 
initial and final states
for typical energies and degrees of polarization of the ILC and CLIC projects.
Numerical results are obtained with the help of Monte Carlo tools: the {\tt ReneSANCe} event generator
and the {\tt MCSANC} integrator.
\end{abstract}

\section{Introduction
\label{sec:introduction}}

At a future high-energy $e^+e^-$ collider, top quarks will be primarily produced via the electroweak annihilation 
process $e^+e^-\to\gamma,Z\to t\bar{t}$. 
The mass of the top quark can then be directly measured  with  a high precision
unreachable at hadron colliders. Looking for effects of new physics in interactions of top quarks is also a very
attractive and valuable objective for future experiments. So, having accurate predictions for various observables
for processes involving top quarks is crucial both for tests of the Standard Model and for new physics searches.

The physical programs for experiments with polarized $e^+$ and $e^-$ beams at
ILC~\cite{ILC:2013jhg,Fujii:2015jha,ILC:2019gyn} and CLIC~\cite{CLICdp:2018esa}
suggest measurement of not only the total cross section for $t\bar{t}$ production but also
different types of asymmetries.
Both the photon and $Z$ boson couplings of the top quark
can be unambiguously measured using these observables~\cite{Englert:2017dev}.

In addition to ILC and CLIC, a scenario 
of longitudinally polarized colliding beams for the  CEPC is considered~\cite{Duan:2023lyp}. 
In particular, these arguments suggest that polarization should be taken into account in 
the corresponding theoretical support and Monte Carlo codes.

Recently
        the study for  the expected precision of the top quark mass and width  in $t \bar{t}$ production using an energy scan around the threshold based on the CEPC scenario,
        assuming
a total integrated luminosity of 100 $fb^{-1}$,  
shows that CEPC is capable of measuring the top quark mass with a precision below 34 MeV \cite{Li:2022iav}.
This study is performed with the help of {\tt QQbar$\_$threshold} package~\cite{Beneke:2016kkb}.

The theoretical uncertainty
for observables of top quark pair production
at the one-loop level ${\cal{O}(\alpha)}$ were estimated for the first time in~\cite{Fujimoto:1987hu}
for the unpolarized case and  in~\cite{Khiem:2015ofa,NhiMUQuach:2017wtf} for different beam polarizations.
Those studies were carried out using the {\tt GRACE-Loop} 
system~\cite{Belanger:2003sd,Yuasa:1999rg}. 

Within the {\tt SANC} project 
we have a library for electroweak (EW) building blocks (self-energies, vertices, boxes)  
in the unitary and $R_\xi$ gauges for the process $e^+e^- \to t \bar{t}$ at the one-loop level
\cite{Andonov:2002rr,Andonov:2002xc}.
We use spin and helicity analysis in combination with the
spinor-helicity formalism to calculate the helicity amplitudes of the one-loop cross section components
\cite{Bondarenko:2020hhn}.

In this paper we consider  theoretical uncertainties associated with electroweak and higher-order effects  
taking into account polarization of the initial and final particles for 
the processes of electron-positron annihilation into a 
top quark pair
\begin{align}
e^+(p_1, \chi_1) + e^-(p_2, \chi_2) 
\to {t}(p_3, \chi_3) + \bar{t}(p_4, \chi_4) (+{\gamma}(p_5, \chi_5)),
\label{eett} 
\end{align}
with arbitrary particle helicities $\chi_i$.
The main goal of this work is to calculate and study three main types of observables
in this process: the total and differential cross section $\sigma_t$, 
several top quark asymmetries, and polarization $P_t$ of the final top quark.
We take a close look at the size of various sources of EW radiation corrections
and carefully examine the QED initial state radiation (ISR) effects. 
  
We consider the beam energies that correspond to the experimental programs of the top quark property studies.
First, at the production threshold, e.g., at 350-GeV center-of-mass (c.m.s.) energy, 
the top quark mass can be measured with a high precision hopefully below 0.1\%.
Second, at 500-GeV c.m.s. energy it is convenient to measure weak and electromagnetic couplings
of the top quark. This energy region also provides an excellent sensitivity to the effect of physics beyond the Standard Model
\cite{Devetak:2010na,
Amjad:2013tlv,Amjad:2015mma}.

QCD radiative corrections to the process of top quark pair production have been extensively studied both at the threshold energy where resummation of higher-order effects is important~\cite{Hoang:2010gu,Hoang:2013uda,Beneke:2015kwa} and above it within pertubative QCD~\cite{Kiyo:2009gb,Gao:2014nva}. The NNLO QCD corrections were also
calculated for unpolarized and polarized forward-backward asymmetries in this process~\cite{Gao:2014eea,Bernreuther:2023ulo}.
In~\cite{Beneke:2017rdn} NNLO electroweak corrections were considered together with QCD effects at the threshold. 
Recently, NLO QCD corrections have also been presented for the process with subsequent decays of the produced (off-shell) top quarks into bottom quarks and $W$ bosons~\cite{Denner:2023grl}. We will no more discuss QCD effects in this paper, leaving the question about their interplay with EW effects for further studies.

The article is organized as follows. The next section contains preliminary remarks and the general notations. 
In Sect. III, we present the numerical results and a comprehensive comparison
of independent MC codes for cross-checking and 
the evaluation of theoretical uncertainties
for observables for polarized and unpolarized cases.
The last section contains a discussion and conclusions.

\section{Radiative corrections to top quark pair production in {\tt SANC}}

We have presented a detailed review of the techniques and results of analytic calculations 
of the NLO EW  scalar form factors and helicity amplitudes of the general 
$e^+e^- \to f \bar{f}$ in
our paper on the $s$-channel lepton-pair production~\cite{Bondarenko:2020hhn}
(note the additional color factor in the final state).

We evaluate the Born level (leading order, LO) cross section $\sigma^{\rm Born}$  contribution 
with both photon and $Z$ boson exchange. 

Gauge invariant subsets of one-loop QED corrections are evaluated separately, 
i.e., the initial state radiation, 
the final state radiation (FSR), 
and the initial-final interference (IFI).

We define
the pure weak contribution as the difference between the complete one-loop electroweak
correction and the pure QED part of it.
The corresponding relative
contributions of the weak and leading higher-order (ho) corrections
will be further denoted as 
$\delta^{\rm weak}$ and $\delta^{\rm ho}$.
The complete one-loop $\delta^{\rm weak}$ consists of pure weak interaction and vacuum polarization (VP) contributions.

We evaluate the leading higher-order EW corrections $\delta^{\rm ho}$ 
to four-fermion processes through the $\Delta\alpha$ and $\Delta\rho$ parameters.
A detailed description of our implementation of this contribution was 
presented in~\cite{Arbuzov:2021oxs}.  
At two-loop level the above corrections consist of the EW at ${\cal{O}}(G_{\mu}^2)$ and the mixed EW$\otimes$QCD at ${\cal{O}}(G_{\mu}\alpha_s)$ parts.

Thus the total EW cross section can be presented as
\begin{eqnarray}
\label{sigma1}
\sigma = 
  \sigma^{\mathrm{Born}}
+ \sigma^{\mathrm{QED}} 
+ \sigma^{\mathrm{weak}} 
+ \sigma^{\mathrm{ho}}.
\end{eqnarray}

Additionally we estimate the multiple photon initial state radiation corrections.
The implementation in {\tt SANC} of these type of corrections in the leading logarithmic approximation (LLA) through the approach of QED structure 
functions~\cite{Kuraev:1985hb,Nicrosini:1986sm}
was described in detail in~\cite{Arbuzov:2021zjc}.
The results are shown up to $\mathcal{O}(\alpha^3L^3)$
finite terms for the exponentiated representation and up to $\mathcal{O}(\alpha^4L^4)$ 
for the order-by-order calculations.
The corresponding relative corrections are denoted below as
$\delta^{\rm LLA,ISR}$. 
The master formula for a general $e^+e^-$ annihilation cross section with ISR QED corrections in the leading logarithmic approximation
has the same structure as the one for the Drell-Yan process.
For ISR corrections in the annihilation channel, the large 
logarithm is $L=\ln({s}/{m_e^2})$, where the total c.m.s.
energy $\sqrt{s}$ is chosen as a factorization scale.
In the LLA approximation, we separate the pure photonic corrections
(marked ``$\gamma$'') 
and the remaining ones, which include the pure pair and mixed photon-pair
effects (marked ``$e^+e^-$'' or ``$\mu^+\mu^-$''). 

The complete two-loop corrections due to initial state radiation
for the unpolarized process $e^+e^-\to\gamma^*,Z$ were first calculated in~\cite{Berends:1987ab}.
Those results were verified and partially corrected in~\cite{Blumlein:2011mi}.
Leading and next-to-leading multiple photon initial state radiation corrections
were computed within the QED structure function formalism in~\cite{Blumlein:2022mrp}
up to the $\mathcal{O}(\alpha^6L^5)$ order, where $L=\ln{s/m_e^2}$ is the so-called large logarithm.

\section{Numerical results and comparisons}
\label{sec2}

Numerical results for
the polarized top quark pair production 
contain estimates of the total cross sections,
as well as energy/angular distributions, 
various polarization effects and 
the study of different types of asymmetries for polarized 
initial and final states.

Here we used the following  set of input parameters:
\begin{eqnarray}
\alpha^{-1}(0) &=& 137.035999084,
\\
M_W &=& 80.379 \; \gev, \quad M_Z = 91.1876 \; \gev, \quad M_H = 125 \; \gev,
\nonumber\\
\Gamma_Z &=& 2.4952 \; \gev, \quad m_e = 0.51099895 \; \mev,
\nonumber\\
m_\mu &=& 0.1056583745 \; \gev, \quad m_\tau = 1.77686 \; \gev,
\nonumber\\
m_d &=& 0.083 \; \gev, \quad m_s = 0.215 \; \gev,
\nonumber\\
m_b &=& 4.7 \; \gev, \quad m_u = 0.062 \; \gev,
\nonumber\\
m_c &=& 1.5 \; \gev, \quad m_t = 172.76 \; \gev.
\nonumber
\end{eqnarray}

The following angular cuts are applied:
\begin{align}
\label{cuts}
|\cos{\vartheta_{t}}| < 0.9,\quad |\cos{\vartheta_{\bar{t}}}| < 0.9,
\end{align}
where $\vartheta_{t}$ and $\vartheta_{\bar{t}}$ are the angles with respect to the electron beam axis.

The results are obtained for the c.m.s. energies $\sqrt{s}=350$ and $500$~GeV
and for unpolarized $(P_{e^+}, P_{e^-}) = (0,0)$,
fully
$(P_{e^+}, P_{e^-}) = (+1,-1),(-1,+1)$
and partially 
$(P_{e^+}, P_{e^-}) = (-0.3,0.8),(0.3,-0.8),(0,0.8),(0,-0.8)$
polarized positron/electron beams.

Most calculations are done in the $\alpha(0)$ EW scheme in order
to have direct access to the effect of vacuum polarization. 
In this scheme, the fine structure constant $\alpha(0)$ and all particle masses are input parameters. 
Additional investigations are performed for scheme dependencies between $\alpha(0)$ and $G_\mu$ EW schemes.

\subsection{Comparison with other codes}

We calculated polarized cross sections at the tree level
for the Born and hard photon bremsstrahlung 
and compared them with the results of the {\tt CalcHEP} \cite{Belyaev:2012qa} 
and {\tt WHIZARD}~\cite{Ohl:2006ae,Kilian:2007gr,Kilian:2018onl}
codes. The Born results agree in all digits for all three codes, and therefore the corresponding table is omitted.

The comparison of the hard bremsstrahlung results is shown in Table~\ref{Table:hard_350500}.
The calculations are done in the $\alpha(0)$ EW scheme
with fixed $100 \%$ polarized initial states 
for $\sqrt{s}=350$ and $500$~GeV,
angular cuts (\ref{cuts})
and an additional cut on the
photon energy  $E_\gamma \ge \omega = 10^{-4} \sqrt{s}/2$.
The table shows results for the unpolarized and fully 
polarized components $(+1,-1)$, $(-1,+1)$, while results for the components $(+1,+1)$, $(-1,-1)$  
are of a different (smaller) order of magnitude, i.e., $1.8(1)\times 10^{-7}$ pb for $\sqrt{s}=350$~GeV
and  $0.238(1)\times 10^{-3}$ pb  for $\sqrt{s}=500$~GeV for all codes.
A very good agreement within statistical errors with the above-mentioned codes is found.

\begin{table}[ht]
\caption{\label{Table:hard_350500}
The tuned triple comparison
of the hard photon bremsstrahlung
        cross section $\sigma^{\text{hard}}$ (pb) 
        between {\tt SANC} (S), {\tt CalcHEP} (C) and {\tt WHIZARD} (W).
}
\begin{center}
\begin{tabular}{lccc}
\hline\hline
$P_{e^+}, P_{e^-}$  & 0,0 &  $+1,-1$ & $-1,+1$ \\
\hline
  \multicolumn{4}{c}
{$\sqrt{s} = $ 350 GeV}\\
S    & 0.13284(1) 
                          & 0.38126(1) & 0.15013(1) \\
W & 0.13282(2)  & 0.38120(1) & 0.15021(5) \\
C & 0.13285(1)  
                          & 0.38124(4) & 0.15014(1) \\
 \multicolumn{4}{c}
 {$\sqrt{s} = $ 500 GeV}\\
S & 0.46733(1) 
    & 1.3090(1) & 0.55987(2) \\
W & 0.46730(2) & 1.3093(4) & 0.55989(4) \\
C & 0.46728(3) 
& 1.3088(1) & 0.55983(5) \\
\hline\hline
\end{tabular}
\end{center}
\end{table}

A comprehensive comparison has been made 
for complete one-loop electroweak radiative corrections
obtained with our codes ({\tt ZFITTER} and {\tt SANC})~\cite{Bardin:2000kn,Andonov:2002rr} 
as well as with the results of the {\tt topfit} code~\cite{Fleischer:2002nn,Fleischer:2003kk}.
We also compared the results of the NLO EW relative corrections calculations
of the {\tt Grace-Loop} code
as a function of the energy for the unpolarized and polarized
cases presented in~\cite{Belanger:2003sd,Yuasa:1999rg}. The qualitative
analysis shows a good agreement.

\subsection{Total cross section}

The corresponding results for the total  cross section (\ref{sigma1}) are presented 
in Tables~\ref{Table:total-unpol}-\ref{Table:total-partpol}
where the relative corrections $\delta^i$ are computed as the ratios (in percent)
of the corresponding 
RC contributions to the Born level cross section. 

\begin{table}[ht]
\caption{\label{Table:total-unpol}
Integrated Born and one-loop cross sections and
relative corrections 
for unpolarized and fully polarized initial beams 
at the c.m.s. energies $\sqrt{s}=350$ and~500~GeV.
}
\begin{center}
\begin{tabular}{lccc}
\hline\hline
$P_{e^+}, P_{e^-}$           & 0,0         &$-1,+1$    & $+1,-1$   \\
\hline 
\multicolumn{4}{c}{$\sqrt{s} = 350$~GeV}\\
$\sigma^{\rm Born}$, pb      & 0.22431(1)  &0.25357(1)   & 0.64367(1)\\
$\sigma^{\rm NLO}$, pb       & 0.16623(1)  &0.20520(1)   & 0.45972(1)\\
$\delta^{\rm NLO}$, \%       & $-25.90(1)$ & $-19.07(1)$ & $-28.58(1)$\\
$\delta^{\rm QED}$, \%       & $-39.87(1)$ & $-40.03(1)$ & $-39.79(1)$\\
$\delta^{\rm VP}$, \%        & 12.84(1)    & 17.51(1)    & 11.00(1)  \\
$\delta^{\rm weak-VP}$, \%   & 1.11(1)     & 3.43(1)     & 0.20(1)  \\
$\delta^{\rm ho}$, \%        & 1.50(1)     & 1.55(1)     & 1.47(1)  \\
\multicolumn{4}{c}{$\sqrt{s} = 500$~GeV}\\
$\sigma^{\rm Born}$, pb      & 0.45030(1) &0.54028(1)&1.2609(1)\\
$\sigma^{\rm NLO}$, pb       & 0.45865(1) &0.60072(1)&1.2334(1)\\
$\delta^{\rm NLO}$, \%       &  1.86(1)   &11.12(1)  & $-2.18(1)$\\
$\delta^{\rm QED}$, \%       & $-4.08(1)$ &$-4.56(1)$ &$-3.91(1)$\\
$\delta^{\rm VP}$, \%        & 12.58(1)   & 16.33(1) & 10.97(1)\\
$\delta^{\rm weak-VP}$, \%   & $-6.63(1)$ & $-5.63(1)$ & $-9.24(1)$\\
$\delta^{\rm ho}$, \%        & 1.73(1)    & 1.82(1)  & 1.69(1) \\
\hline\hline
\end{tabular}
\end{center}
\end{table}

\begin{table}[ht]
\caption{\label{Table:total-partpol}
Integrated Born and one-loop cross sections and
relative corrections
for partially polarized initial beams 
at the c.m.s. energies $\sqrt{s}=350$ and~500~GeV.
}
\begin{center}
\begin{tabular}{lcccc}
\hline\hline
$P_{e^+}, P_{e^-}$           & $0.3,-0.8$ & $-0.3,0.8$ & $0,-0.8$  & $0,0.8$\\
\hline 
\multicolumn{5}{c}{$\sqrt{s} = 350$~GeV}\\
$\sigma^{\rm Born}$, pb      &  0.38542(1) & 0.17086(1)  & 0.30232(1)  & 0.14629(1)\\
$\sigma^{\rm NLO}$, pb       &  0.27612(1) & 0.13612(1)  & 0.21713(1)  & 0.11532(1)\\
$\delta^{\rm NLO}$, \%       & $-28.36(1)$ & $-20.33(1)$ & $-28.18(1)$ & $-21.17(1)$\\
$\delta^{\rm QED}$, \%       & $-39.80(1)$ & $-40.01(1)$ & $-39.81(1)$ & $-39.99(1)$ \\
$\delta^{\rm VP}$, \%        &  11.15(1)   & 16.65(1)    & 11.28(1)    & 16.08(1)\\
$\delta^{\rm weak-VP}$, \%   &  0.27(1)    & 3.01(1)     & 0.33(1)     & 2.72(1)\\
$\delta^{\rm ho}$, \%        &  1.48(1)    & 1.54(1)     & 1.48(1)     & 1.53(1)\\
\multicolumn{5}{c}{$\sqrt{s} = 500$~GeV}\\
$\sigma^{\rm Born}$, pb      & 0.75654(1) & 0.36020(1) & 0.59444(1) & 0.30617(1)\\
$\sigma^{\rm NLO}$, pb       & 0.74267(1) & 0.39468(4) & 0.58522(1) & 0.33212(1)\\
$\delta^{\rm NLO}$, \%       & $-1.83(1)$ & 9.58(1)    & $-1.55(1)$ & 8.48(1)\\
$\delta^{\rm QED}$, \%       & $-3.92(1)$ & $-4.46(1)$ & $-3.91(1)$ & $-4.40(1)$ \\
$\delta^{\rm VP}$, \%        & 11.11(1)   & 15.67(1)   & 11.22(1)   & 15.22(1)\\
$\delta^{\rm weak-VP}$, \%   & $-9.02(1)$ & $-1.63(1)$ & $-8.84(1)$ & $-2.35(1)$ \\
$\delta^{\rm ho}$, \%        & 1.69(1)    & 1.80(1)    & 1.69(1)    & 1.79(1)\\
\hline\hline
\end{tabular}
\end{center}
\end{table}

One-loop and ho weak-interaction corrections strongly depend on the choice of the EW scheme,
and the total weak corrections in the $G_\mu$ scheme are smaller by about
5-6\% than in the $\alpha(0)$ one.

The integrated cross sections for the weak and leading higher-order corrections in the $\alpha(0)$ and $G_\mu$ schemes and their relative difference
\begin{eqnarray}
\delta_{G_\mu/\alpha(0)} = \frac{\sigma_{G_\mu}}{\sigma_{\alpha(0)}} - 1,\, \%
\label{rgmual0}
\end{eqnarray}
are presented in Table~\ref{Table:delta_350ewscheme}. 
Ratio~(\ref{rgmual0}) 
shows the stabilization of the results and can be considered as
an estimation of the theoretical uncertainty of the weak and h.o. contributions.
As is well known, the difference between two EW schemes in the LO is just the
ratio of the EW couplings and gives $\delta^{\rm LO}_{G_\mu/\alpha(0)} = 7.5\%$. 
As is seen from the Tables, the weak contribution reduces the difference 
to about
2\% at the energy of 350 GeV and 1.5\% at 500 GeV. 
Moreover, the
sum of the weak and ho contributions reduces the difference 
to about 
0.33\% at 350 GeV and $-0.4$\% at 500 GeV.

\begin{table}[ht]
\caption{
\label{Table:delta_350ewscheme}Integrated Born and weak contributions 
to the cross section
and higher-order leading corrections in two EW schemes:
$\alpha(0)$ and $G_\mu$ 
at the c.m.s. energies $\sqrt{s}=350$ and 500~GeV.
}
\begin{center}
\begin{tabular}{lcc}
\hline\hline
$\sqrt{s}$, GeV & 350         & 500\\
\hline
{\sba, pb}      & 0.22431(1)  & 0.45030(1)\\
{\sbg, pb}      & 0.24108(1)  & 0.48398(1)\\
{\dbg, \%}      & 7.48(1)     & 7.48(1)\\
{\swa, pb}      & 0.25564(1)  & 0.47705(1)\\
{\swg, pb}      & 0.26055(1)  & 0.48420(1)\\
{\dwg, \%}      & 1.92(1)     & 1.50(1)\\
{\swha, pb}     & 0.25900(1)  & 0.48483(1)\\ 
{\swhg, pb}     & 0.25986(1)  & 0.48289(1)\\
{\dwhg, \%}     & 0.33(1)     & $-0.40$(1)\\
\hline\hline
\end{tabular}
\end{center}
\end{table}

\subsubsection{\bf Multiple photon ISR corrections in the LLA approximation}

Here we discuss the estimation of the initial-state photon radiations in detail.
In Table~\ref{Table:LLAQED5}  we show  the corresponding results for the 
multiple photon ISR corrections
of different order of ${\cal O}(\alpha^nL^n), n=2-4$ in the LLA approximation for
the c.m.s. energies $\sqrt{s}=350$~{GeV} and $500$~{GeV}
in the $\alpha(0)$ EW scheme.
The relative corrections $\delta^i$ are computed as the ratios (in percent) of the corresponding 
RC contributions to the Born level cross section.
\begin{table}[ht]
\caption{\label{Table:LLAQED5}
Multiple photon ISR relative corrections $\delta$ ($\%$) in the LLA approximation
 at $\sqrt{s} = 350$ and $500$ GeV with cuts~(\ref{cuts}). 
}
\centering
\begin{tabular}{lcc}
\hline\hline
$\sqrt{s}$, GeV                & 350           & 500 \\
\hline
$\mathcal{O}(\alpha L)$, $\gamma$        & $-42.546(1)$ & $-3.927(1)$ \\
$\mathcal{O}(\alpha^2L^2)$, $\gamma$     & $+8.397(1)$ & $-0.429(1)$\\
$\mathcal{O}(\alpha^2L^2)$, $e^+e^-$     & $-0.460(1)$ & $-0.030(1)$  \\
$\mathcal{O}(\alpha^2L^2)$, $\mu^+\mu^-$ & $-0.277(1)$ & $-0.018(1)$ \\
$\mathcal{O}(\alpha^3L^3)$, $\gamma$     & $-0.984(1)$ & $+0.021(1)$\\
$\mathcal{O}(\alpha^3L^3)$, $e^+e^-$     & $+0.182(1)$ & $-0.012(1)$\\
$\mathcal{O}(\alpha^3L^3)$, $\mu^+\mu^-$ & $+0.110(1)$ & $-0.008(1)$ \\
$\mathcal{O}(\alpha^4L^4)$, $\gamma$     & $+0.070(1)$ & $+0.002(1)$\\
\hline\hline
\end{tabular}
\end{table}
The most significant contribution is of course the 
photonic one of the order ${\cal O}(\alpha L)^2$. 
For the c.m.s. energy $\sqrt{s}=350$~GeV, the dominant contributions of the second order are about
$+8.397\%$ for $\gamma$ and $-0.460\%$ for $e^+e^-$-pairs ($-0.277\%$ for $\mu^+\mu^-$-pairs).
Similar behaviour occurs for the energy 
$\sqrt{s}=500$~GeV, but orders of magnitudes of the multiple photon corrections are much smaller.

When considering multiple photon corrections, we see that it is certainly sufficient to take into account corrections 
up to the fourth order.

\subsection{Differential distributions}

\subsubsection{Angular distributions}

\begin{figure}
\begin{center}
\includegraphics[width=0.49\linewidth]{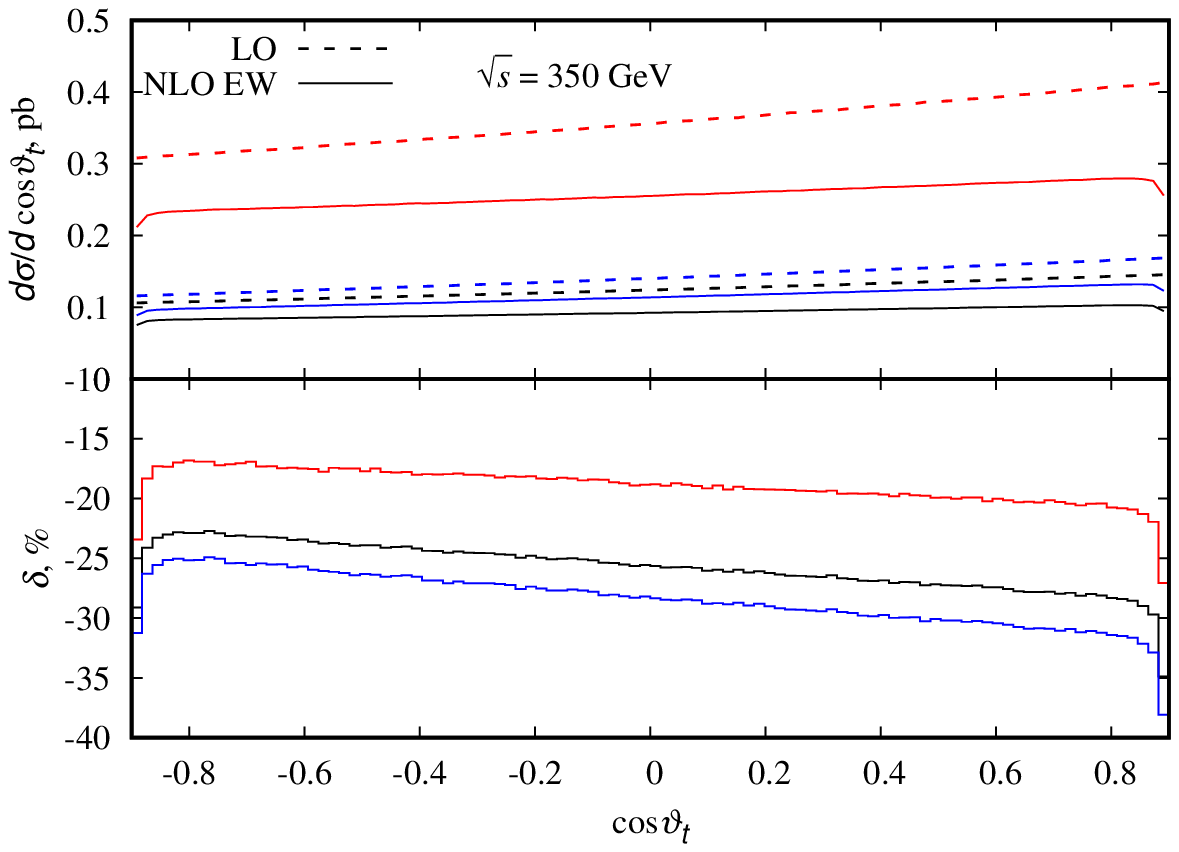}
\includegraphics[width=0.49\linewidth]{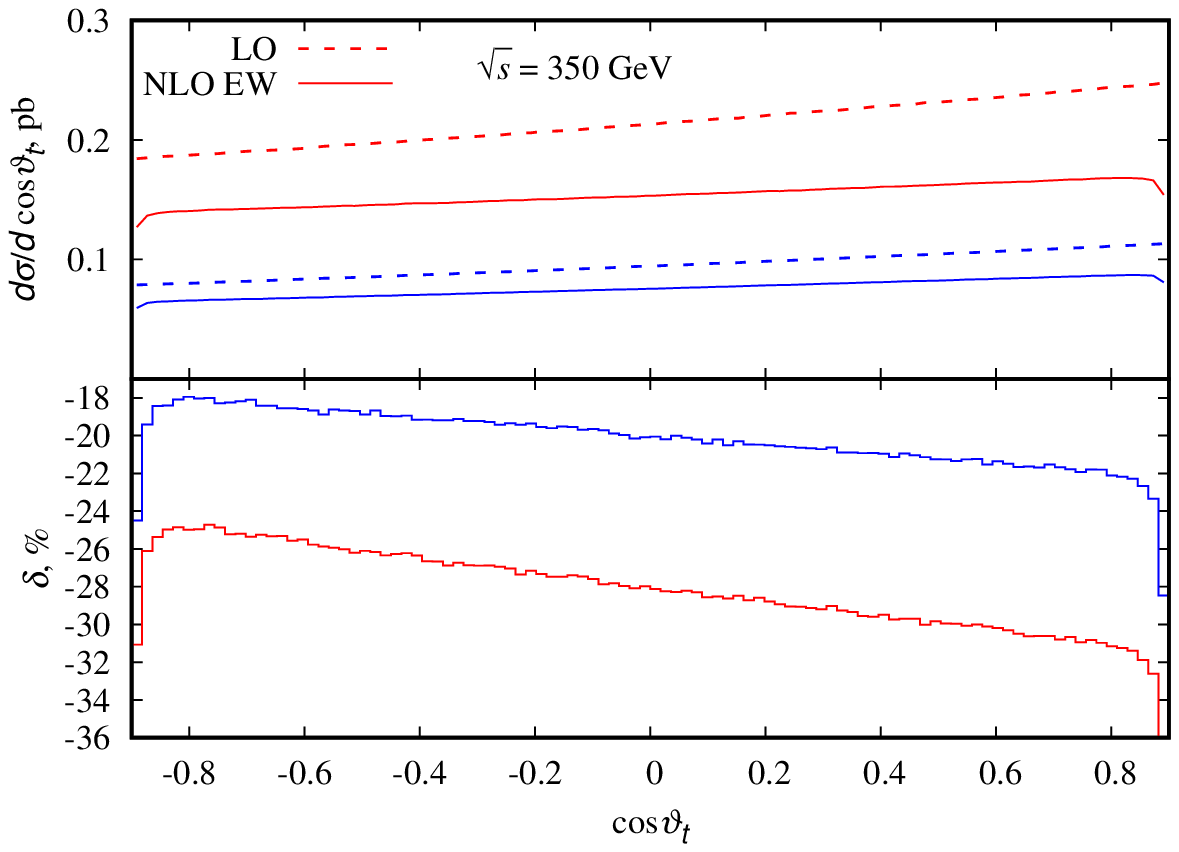}\\
\caption{
{LO and EW NLO cross sections and relative corrections
at $\sqrt{s} = 350$~GeV 
with (un)polarized initial beams.}
}
\label{fig_xcross_350}
\end{center}
\begin{center}
\includegraphics[width=0.49\linewidth]{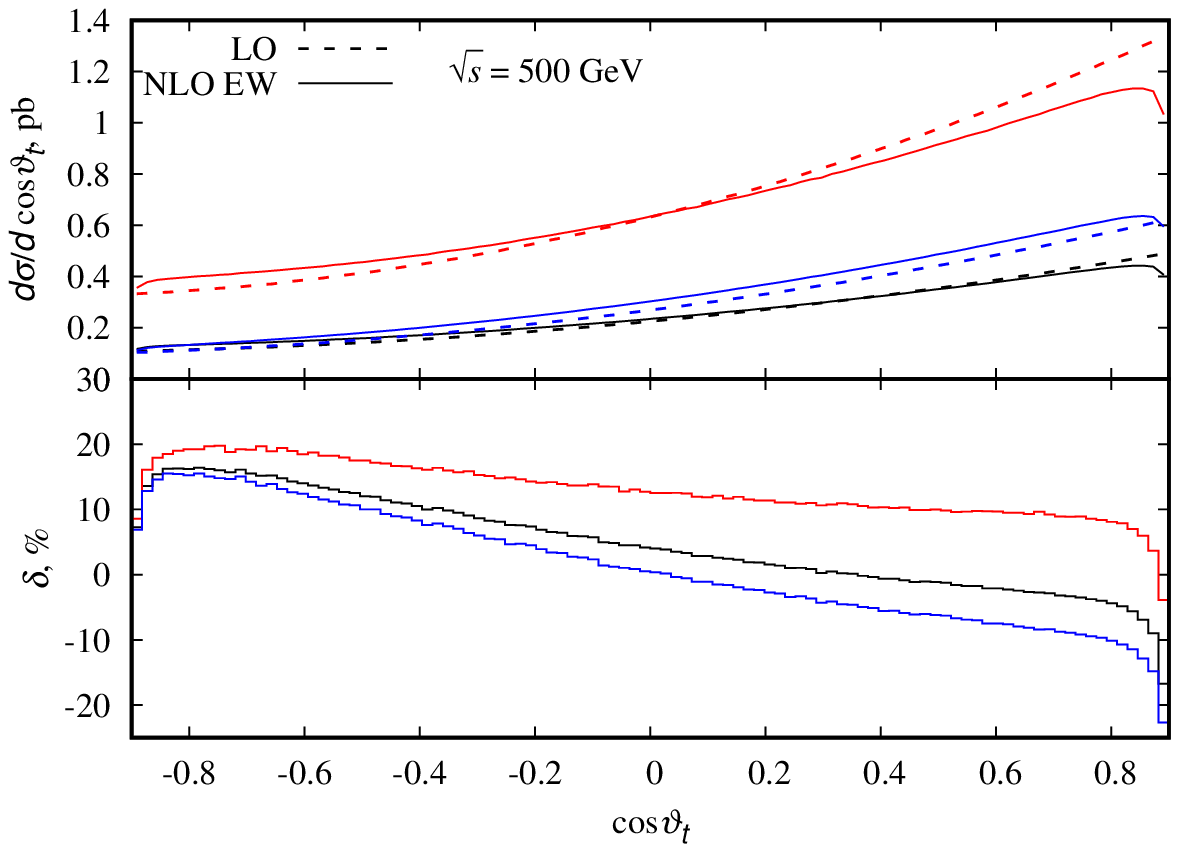}
\includegraphics[width=0.49\linewidth]{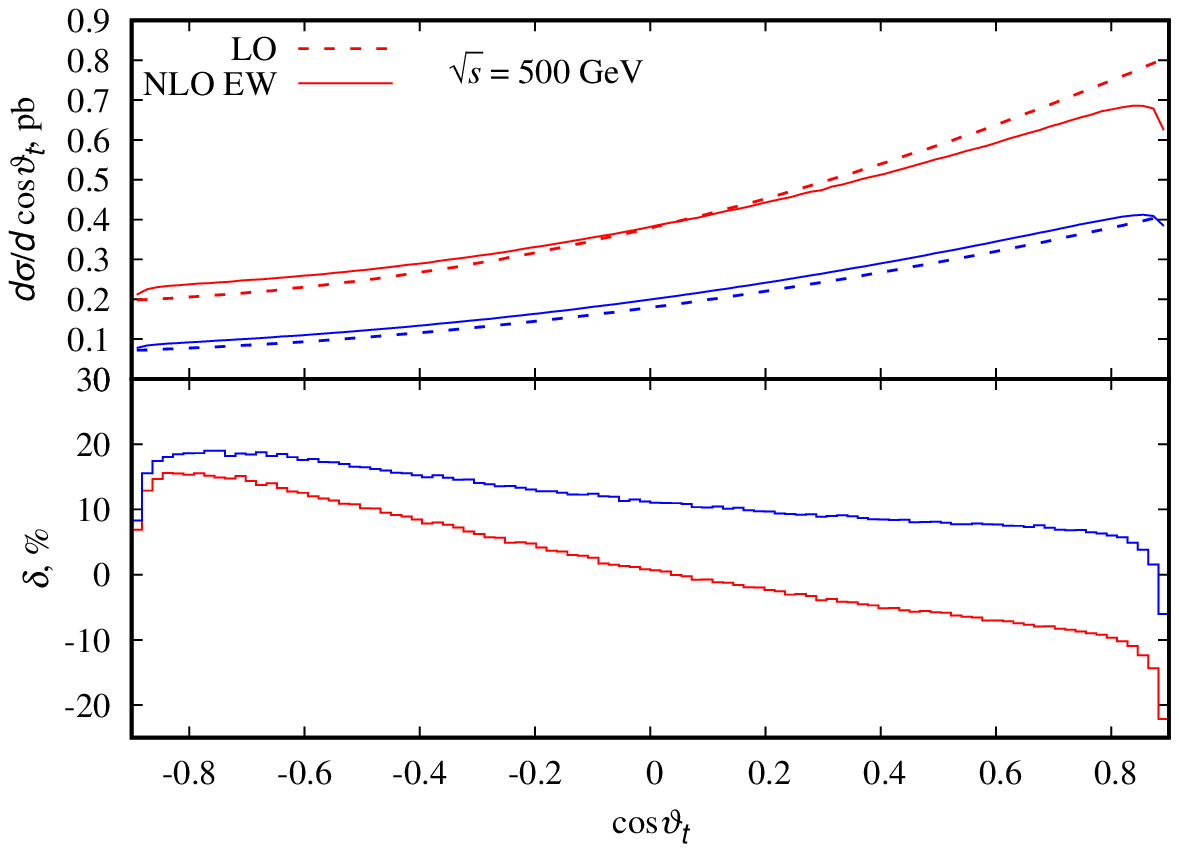}
\caption{
The same as in Fig.~\ref{fig_xcross_350} but for $\sqrt{s} = 500$~GeV.
}
\label{fig_xcross_500}
\end{center}
\end{figure}

In Figs.~\ref{fig_xcross_350}-\ref{fig_xcross_500}
the LO (dashed line) and NLO EW (solid line) cross sections
(upper panel) as well as the relative corrections (lower panel)
are shown.
The left part of Fig.~\ref{fig_xcross_350} corresponds to
the unpolarized (black), and fully polarized,
with $(P_{e^+},P_{e^-}) = (+1,-1)$ (red) and $(-1,+1)$ (blue),
initial beams, while the right one
shows the partially polarized initial beams with
$(P_{e^+},P_{e^-}) = (+0.3,-0.8)$ (red) and $(-0.3,+0.8)$ (blue)
for the energy $\sqrt{s} = $ 350 GeV.
Figure ~\ref{fig_xcross_500} shows the same but for $\sqrt{s} = $ 500 GeV.

The radiative corrections significantly reduce cross sections at the energy $\sqrt{s} = $ 350 GeV
in the whole range of the scattering angles.
The corresponding relative corrections are large, negative and varied 
from $-32 \%$ to $-12 \%$
for unpolarized/fully polarized states. 
The real planned polarized states in the ILC experiment (right panel) show significant
dependence on the polarization of the initial beams, namely,
for  
$(P_{e^+},P_{e^-}) = (+0.3,-0.8)$
the relative corrections are $-(25-32) \%$ while for $(-0.3,+0.8)$ they are $-(18-20) \%$.

At the c.m.s energy $\sqrt{s} = $ 500 GeV the LO and NLO EW differential cross sections can cross each other and
therefore the relative corrections can change the sign. The dependence on polarization is also strong,
and $\delta$ are from $15$ \% to $-10$ \% for $(+0.3,-0.8)$ and from $20$ \% to $0$ \% for $(-0.3,+0.8)$.

It should also be noted that the nonphysical dips in the first and last bins of the relative correction histograms are 
due to the angular limits~\ref{cuts} and can be removed by applying wider cuts.
 
\subsubsection{Energy dependence}

\begin{figure}[ht]
\begin{center}
\includegraphics[width=0.8\textwidth]{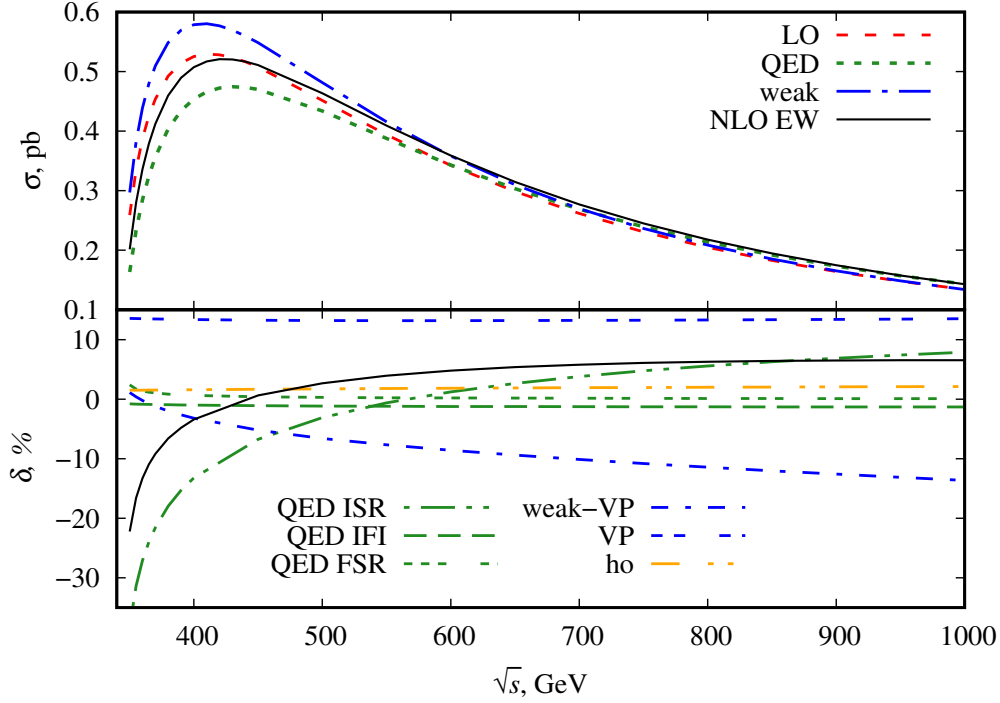}
\end{center}
\caption{
\label{fig:eettdelta}
The LO and NLO EW corrected unpolarized cross sections 
and the relative corrections in parts 
as a function of the c.m.s. energy.
}
\end{figure}

In  Fig.~\ref{fig:eettdelta} the unpolarized cross sections for the LO and for NLO EW in parts are presented.
The upper panel shows the cross sections for the QED and weak gauge invariant contributions to NLO EW while the
lower panel demonstrates the corresponding relative corrections to the Born cross section
subdivided inside the QED (ISR, IFI, FSR) and weak (VP and weak-VP) sectors.
The contributions of the leading higher-order corrections are present as well.

It is seen from the figure that the total NLO EW contribution
near the threshold at the c.m.s energy $\sqrt{s}$ = 350 GeV
is defined by large negative QED (about $-35$ \%) and positive weak 
(15 \%) contributions, then at approximately $\sqrt{s}$ = 450 GeV
they compensate each other, and above that energy QED part dominates.
It should be noted that in the QED contribution the ISR part dominates while in the weak contribution the VP part dominates.
The leading higher-order two-loop contributions
are rather low, about 1.5-2\%, but play an important role in the EW scheme dependency stabilization.

\subsection{Asymmetries}

In this section we analyze the effect of radiative corrections 
for different types of asymmetries:
the left-right $\alr$ and forward-backward $\afb$ asymmetries,
as well as the final state quark polarization $\ptop$. 

\subsubsection{Left-right asymmetry $\alr$} \label{Sect:ALR}

The asymmetry $\alr$ is defined in the following form:
$$
\alr=\frac{\sigma_{\rm LR}-\sigma_{\rm RL}}{\sigma_{\rm LR}+\sigma_{\rm RL}},
$$
where $\sigma_{\rm LR}$  and $\sigma_{\rm RL}$ are the cross sections for 
the fully polarized electron-positron $e^-_{\rm L}e^+_{\rm R}$ and  $e^-_{\rm R}e^+_{\rm L}$ 
initial states, respectively. For the given definition,
$\alr$ does not depend on the degrees of the initial beam polarization, but
this type of asymmetry is sensitive to electroweak interaction effects.

\begin{figure}[!h]
\begin{center}
\includegraphics[width=0.49\linewidth]{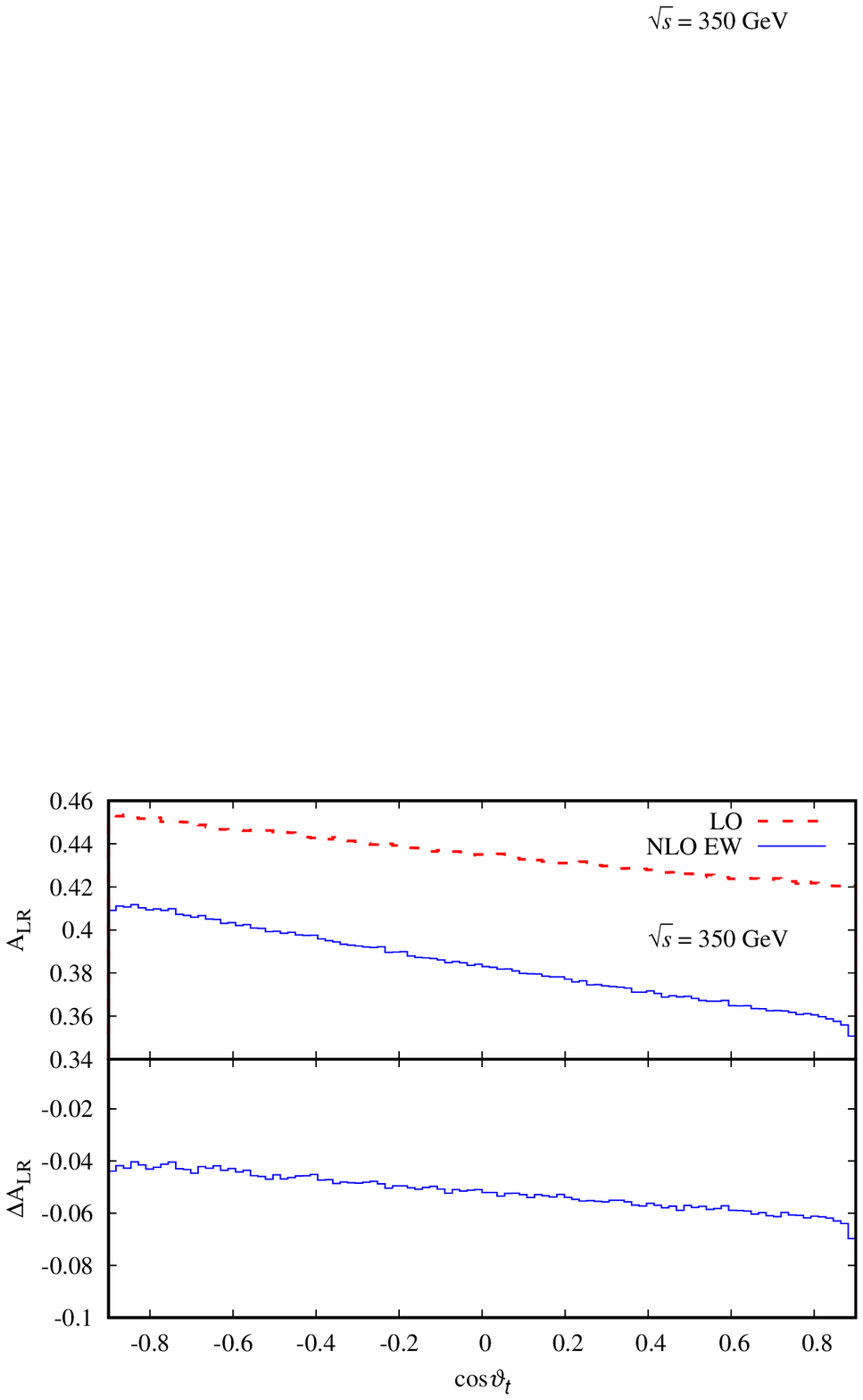}
\includegraphics[width=0.49\linewidth]{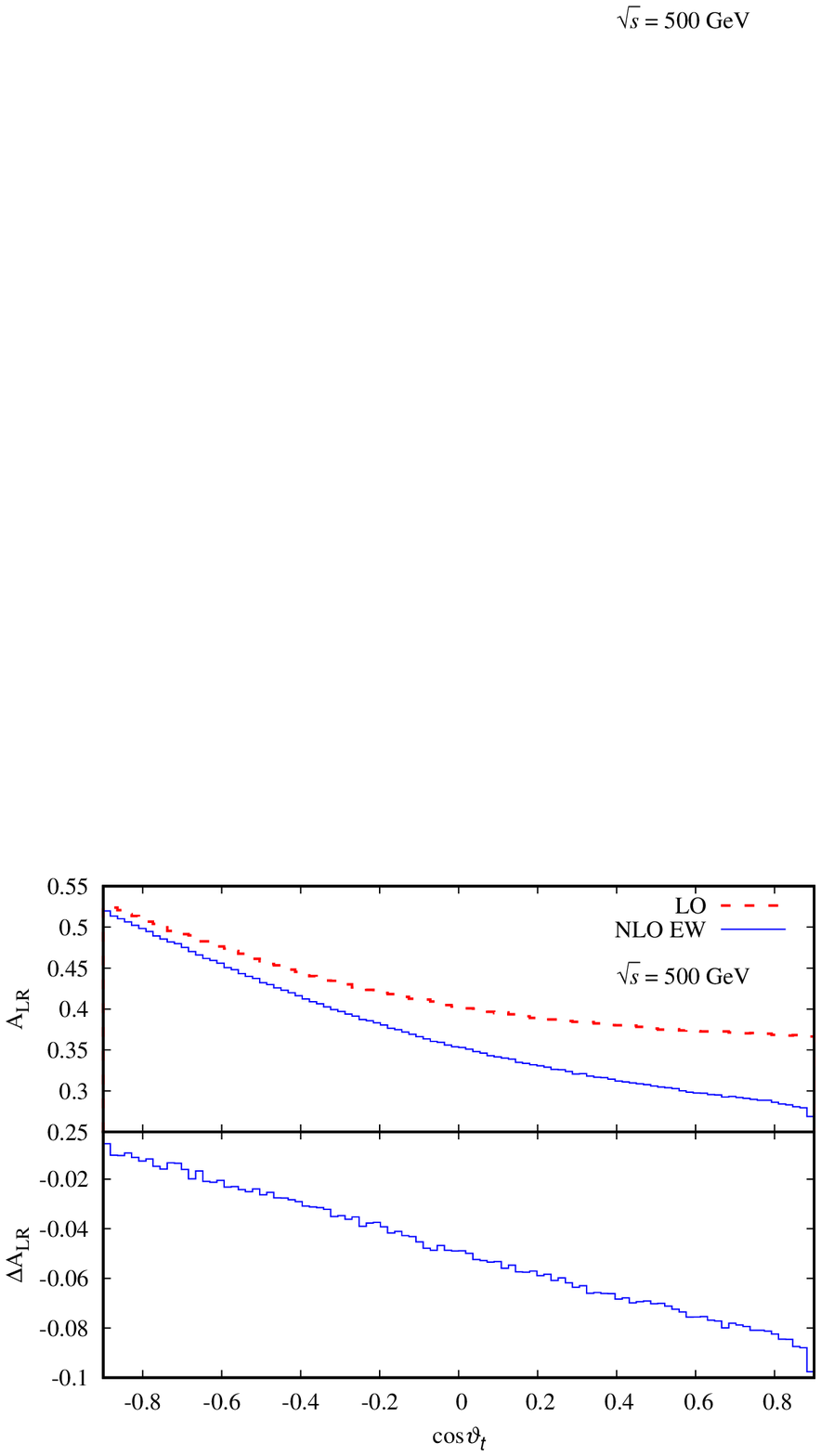}
\caption{
The asymmetry $\alr$  in the Born and one-loop approximations 
at $\sqrt{s} = 350$~GeV (left) and $\sqrt{s} = 500$~GeV (right) 
vs. the cosine of the scattering angle.
}
\label{fig_alr_350500}
\end{center}
\end{figure}

In Fig.~\ref{fig_alr_350500},
the left-right asymmetry distributions 
for the Born and one-loop contributions
are shown as a function of the cosine of the top quark scattering angle.
The corresponding shift of the asymmetry 
$$\Delta \alr = \alr ({\rm NLO\ EW}) - \alr ({\rm LO})$$
is shown in the lower panel.

At the c.m.s. energy $\sqrt{s} = 350$~GeV $\Delta\alr$ changes from about $-0.04$ to $-0.06$ while 
at $\sqrt{s} = 500$~GeV it changes
from about $-0.01$ to $-0.09$ over the whole range of the top quark scattering angles.

\subsubsection{Forward-backward asymmetry $\afb$} \label{Sect:AFB}

The forward-backward asymmetry is defined as
$$
A_{\rm FB} = \frac{\sigma_{\rm F}-\sigma_{\rm B}}{\sigma_{\rm F}+\sigma_{\rm B}},
$$
where
$$
\sigma_{\rm F} = \int\limits_0^1 \frac{d\sigma}{d\cos\vartheta_t}d\cos\vartheta_t,\quad
\sigma_{\rm B} = \int\limits_{-1}^0 \frac{d\sigma}{d\cos\vartheta_t}d\cos\vartheta_f.
$$

\begin{figure}[!h]
\begin{center}
\includegraphics[width=0.49\linewidth]{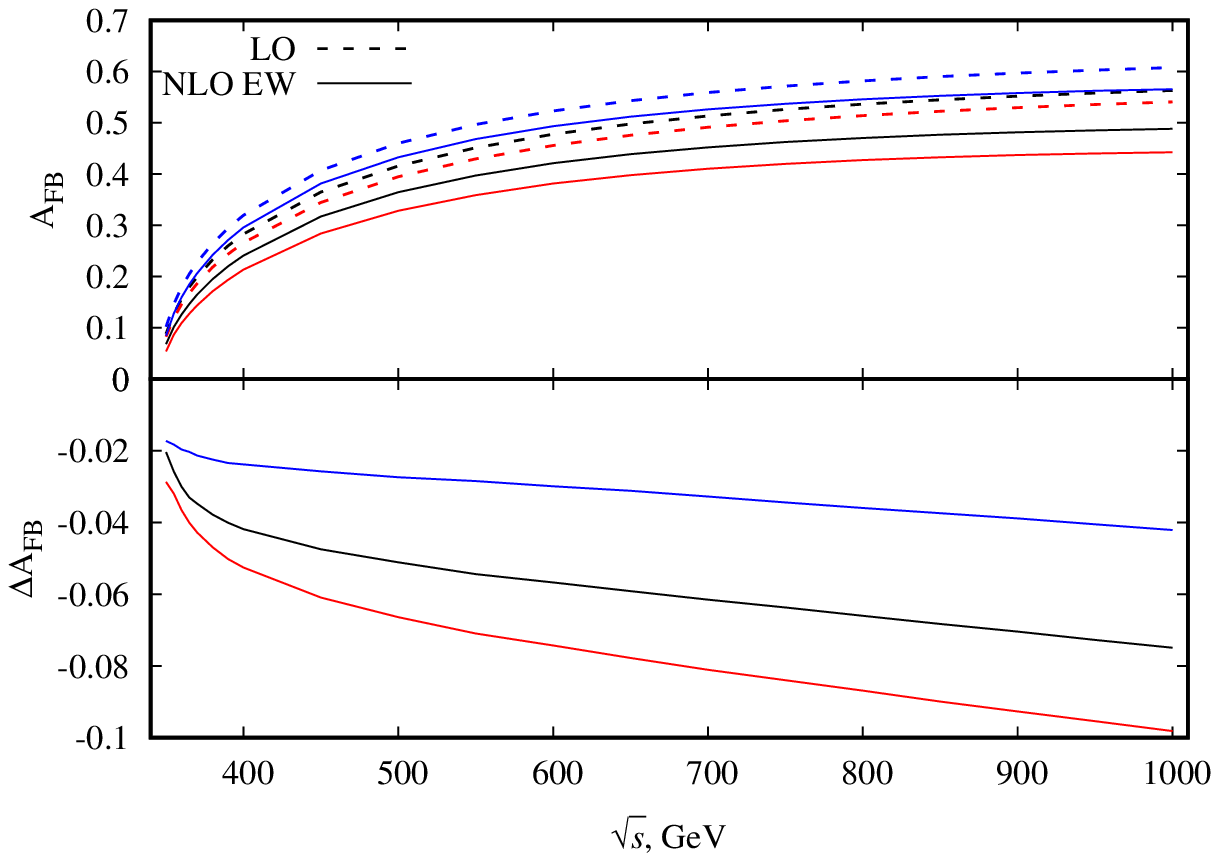}
\includegraphics[width=0.49\linewidth]{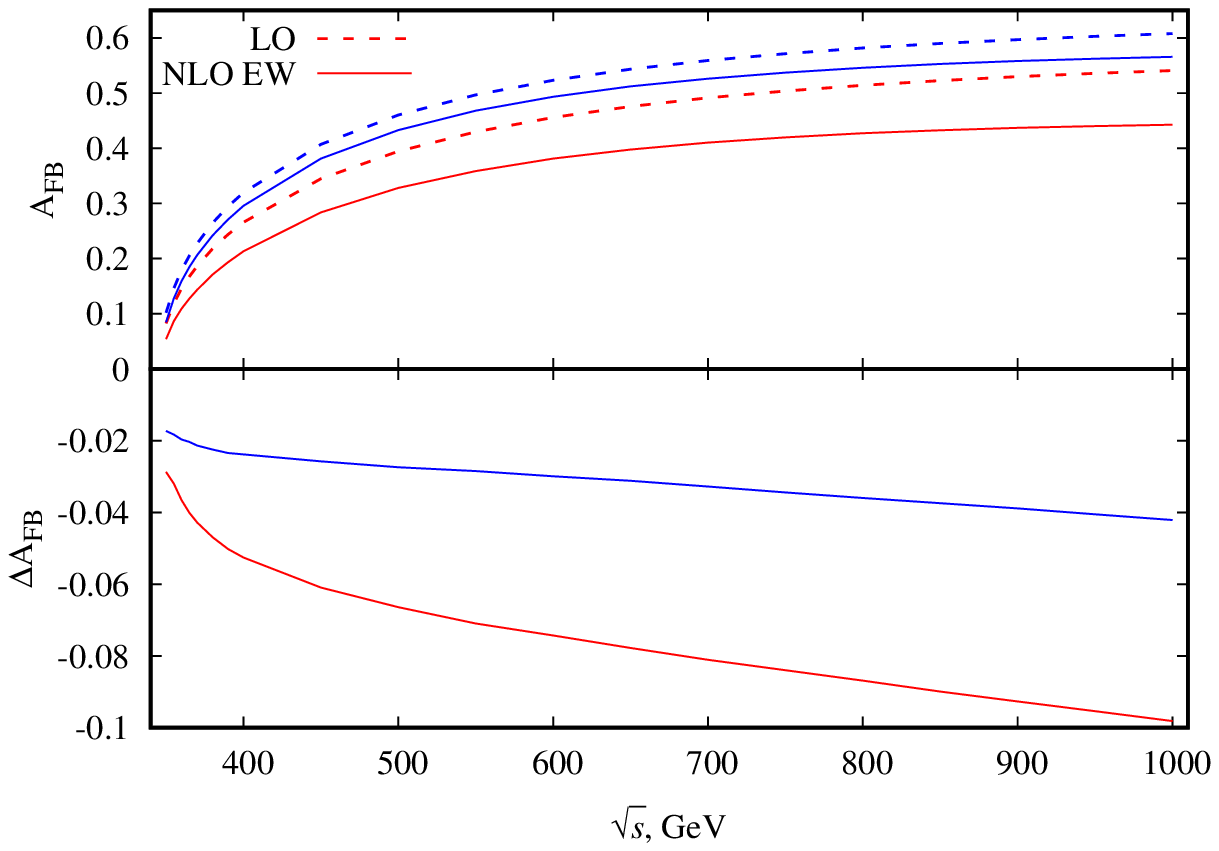}
\caption{
The asymmetry $\afb$  in the Born and one-loop
approximations and the corresponding shift
as a function of the c.m.s. energy. Details are in the text.
}
\label{fig_afb_350500}
\end{center}
\end{figure}

In Fig.~\ref{fig_afb_350500}, the asymmetry $\afb$  
in the Born (dashed) and one-loop (solid) 
approximations (upper panel) and the corresponding shift 
$$\Delta \afb = \afb ({\rm NLO\ EW}) - \afb ({\rm LO})$$
(lower panel) as a function of $\sqrt{s}$ are presented.
On the left, the black lines are for the unpolarized initial beams
while the red and blue ones are for the fully polarized cases $(P_{e^+},P_{e^-}) = (+1,-1)$
and $(-1,+1)$, respectively.
On the right, the red and blue lines are for the partially polarized beams 
with $(P_{e^+},P_{e^-}) = (+0.3,-0.8)$ and
$(-0.3,+0.8)$, respectively.

One can see that a combination of degrees of initial particles polarization  
can either increase $(P_{e^+},P_{e^-}) = (0.3,-0.8)$ or decrease $(P_{e^+},P_{e^-}) = (-0.3,-0.8)$ $\afb$ with respect to the unpolarized case.

The asymmetry $\afb$ is zero both for LO and NLO EW at the threshold 
and increase with increasing energy.
The NLO EW corrections decrease the LO results, and $\Delta \afb$ is always negative in the c.m.s energy range $\sqrt{s}=350-1000$ GeV.

\subsubsection{Final-state fermion polarization $P_t$}  \label{Sect:Ptau}

The polarization of a final-state top quark $P_{t}$ can be expressed as 
the ratio between the difference of the cross sections for the right- and left-handed 
final state helicities and their sum
$$
P_{\rm t} =\frac{\sigma_{\rm R_t}-\sigma_{\rm L_t}}{\sigma_{\rm R_t}+\sigma_{\rm L_t}}.
$$

\begin{figure}[!h]
\begin{center}
\includegraphics[width=0.49\linewidth]{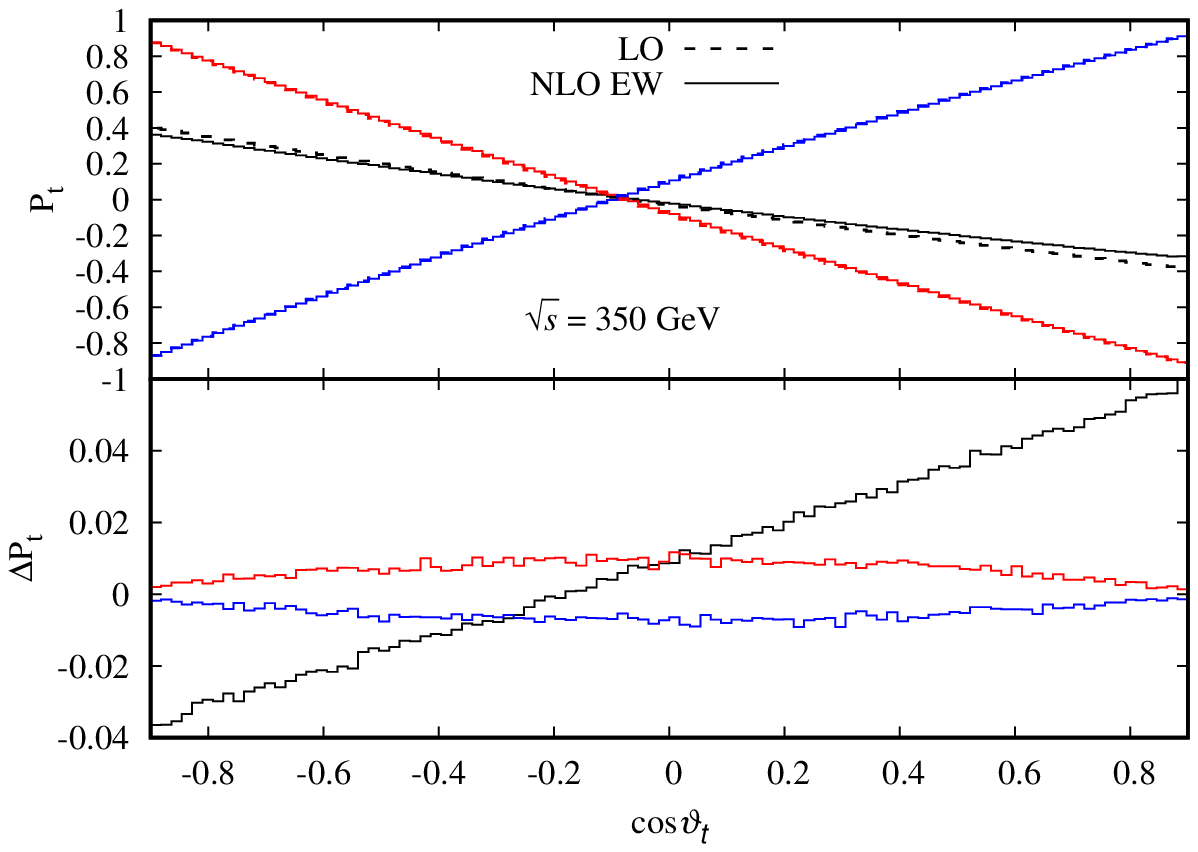}
\includegraphics[width=0.49\linewidth]{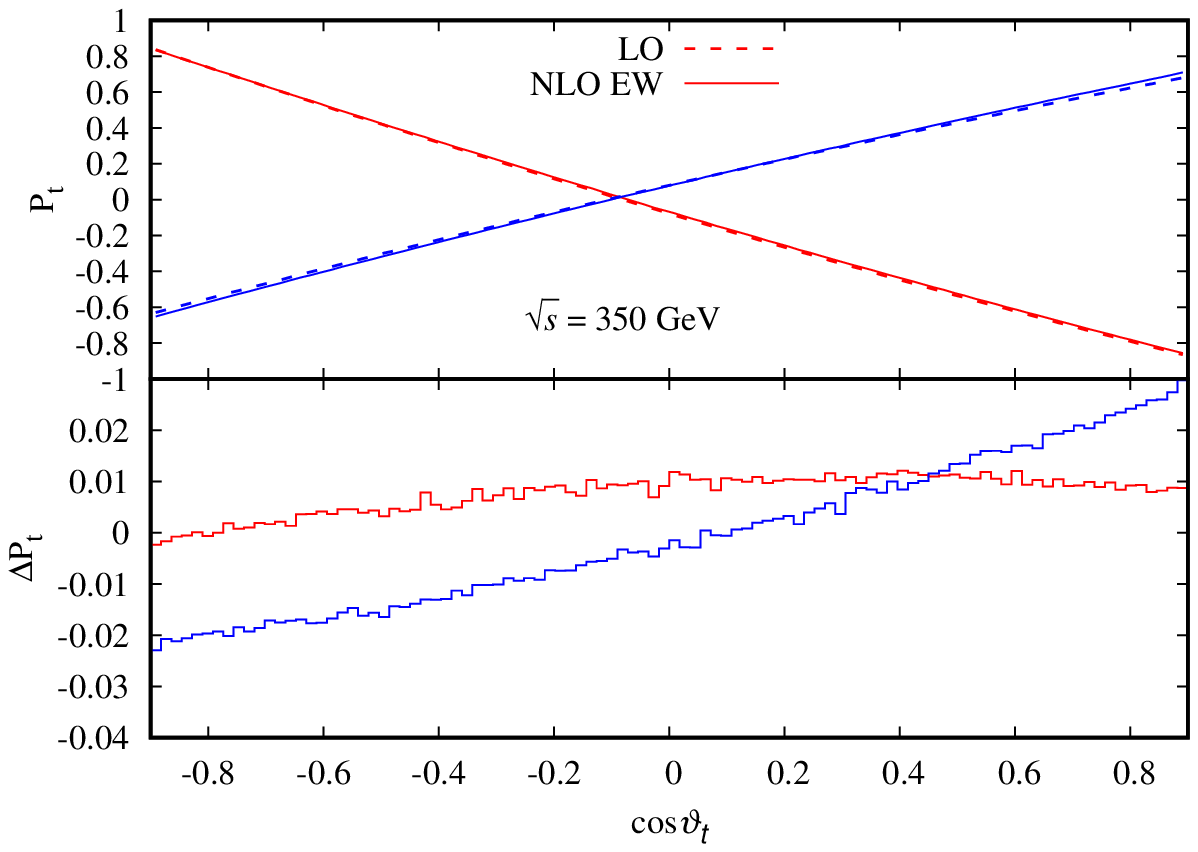}
\caption{
Top quark polarization $P_t$ in the Born and one-loop approximations
and the corresponding shifts $\Delta P_t$ vs. the scattering angle
at the c.m.s. energy $\sqrt{s} = 350$~GeV. 
Details are in the text.
}
\label{fig_pt_350}
\end{center}
\end{figure}

\begin{figure}[!h]
\begin{center}
\includegraphics[width=0.49\linewidth]{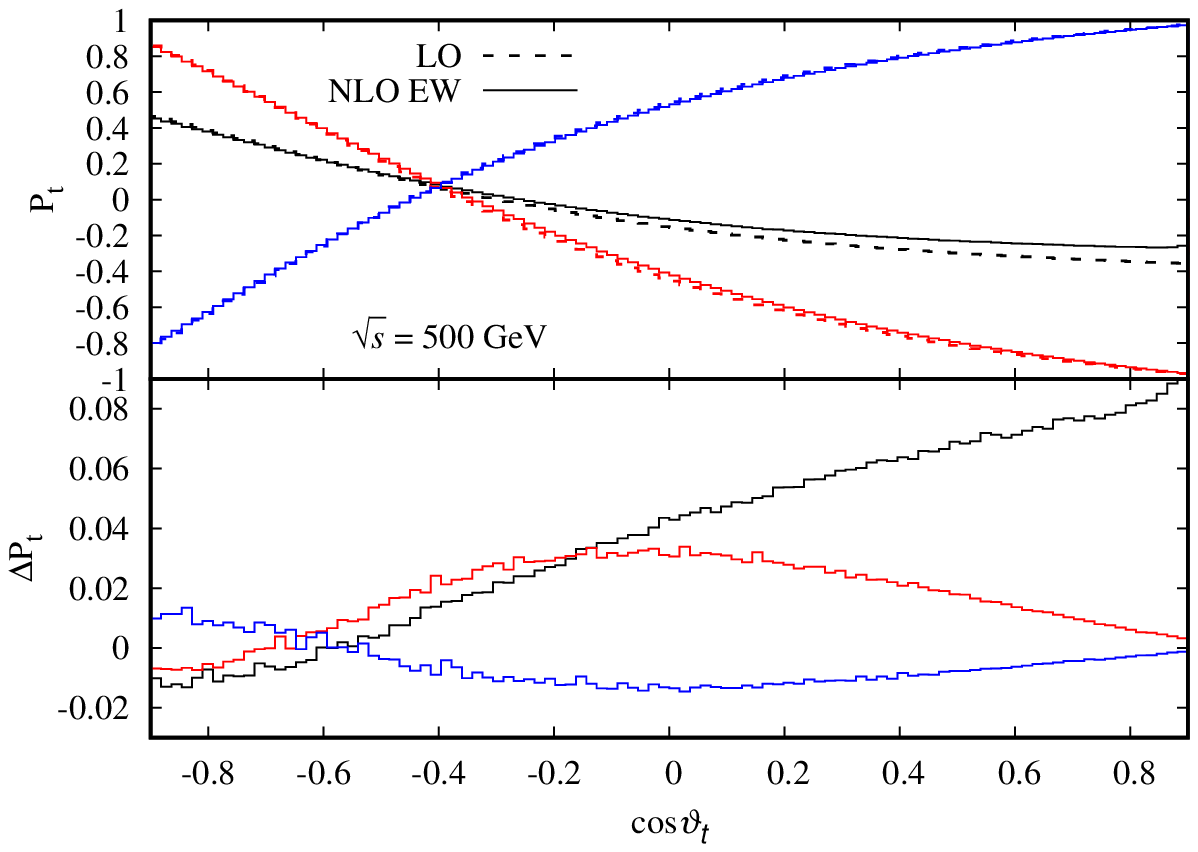}
\includegraphics[width=0.49\linewidth]{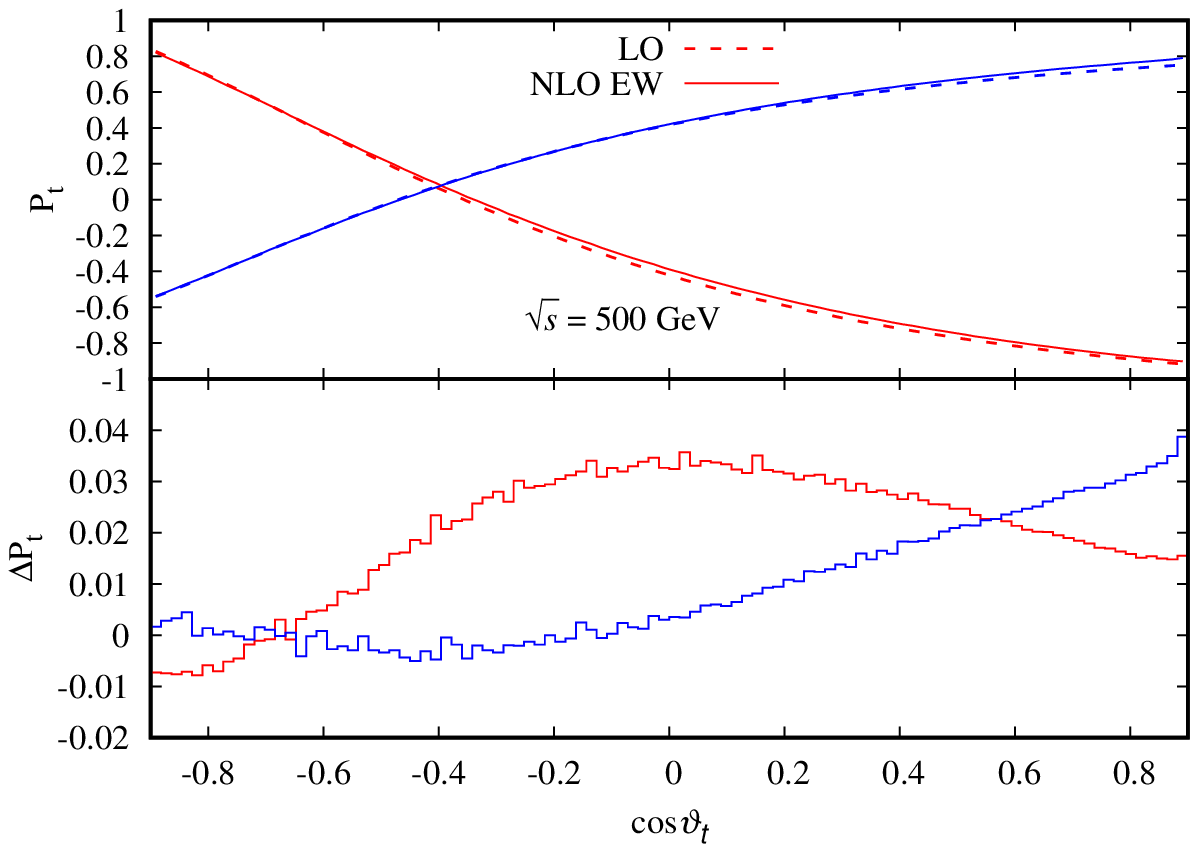}
\caption{
The same as in Fig.~\ref{fig_pt_350} but for $\sqrt{s} = 500$~GeV.
}
\label{fig_pt_500}
\end{center}
\end{figure}

In Fig.~\ref{fig_pt_350}(\ref{fig_pt_500}), the top quark polarization in the Born (dashed) 
and one-loop (solid) approximations (upper panel) and the corresponding shift (lower panel)
$$\Delta P_{t}=P_{t}(\mathrm{NLO\ EW})-P_{t}(\mathrm{LO})$$
at the c.m.s. energy $\sqrt{s} = 350$~(500)~GeV.
On the left, the black lines are for the unpolarized initial beams
while the red and blue ones are for the fully polarized cases of $(P_{e^+},P_{e^-}) = (+1,-1)$ and $(-1,+1)$, respectively.
On the right, the red and blue lines are for partially polarized beams 
with $(P_{e^+},P_{e^-}) = (+0.3,-0.8)$ and
$(-0.3,+0.8)$, respectively.

This asymmetry is important for studying possible
manifestations of CP violation beyond the Standard Model~\cite{Ladinsky:1992vv}.

The results for $P_{\rm t}$ are very much affected by initial beam polarizations.
The difference $\Delta P_{t}$ also depends on the c.m.s. energy and initial beam
polarizations. The largest values of $\Delta P_{t}$  for unpolarized initial beams are $-0.04$ to $0.05$ at $\sqrt{s}=350$~GeV and $-0.01$ to $0.08$ at
500 GeV. Polarization of the initial states significantly reduces $\Delta P_{t}$  
both at 350 and 500~GeV.

\section{Conclusion}
\label{sec:connclusion}

In this paper, we investigated electroweak corrections to the process
of electron-positron annihilation into a top quark pair with allowance for polarizations of the initial and final particles.
Numerical results are presented for energies and polarizations which
are typical of the future CLIC and ILC linear $e^+e^-$ collider projects.

The calculated polarized cross sections at the tree level
for the Born and  hard photon bremsstrahlung were thoroughly
compared with the {\tt CalcHEP} and {\tt WHIZARD} results. 
A very good agreement was observed. 

Then virtual (loop) EW corrections were calculated within the SANC system.
Numerical studies were carried out for several observables 
in the $t \bar{t}$ production process for unpolarized and polarized beams
with taking into account the NLO EW level, higher-order corrections, 
and multiple photon ISR corrections.

We considered a set of benchmark polarizations and found 
that the relative effects of $e^{\pm}$ polarizations on the EW radiative correction 
are quite sizeable. In other words, one can not use the same correction factors
for the cases of different  degrees of beam polarization.
The NLO EW corrections qualitatively agreed with
the {\tt Grace-Loop} results.

Various asymmetries which can be measured in the given process were analyzed.
For all the asymmetries, the NLO EW effects are found to be quite sizable. 
The magnitude of EW radiative corrections to asymmetries at 500~GeV c.m.s. energy 
is higher than at 350~GeV in most cases. 

It was demonstrated that a considerable EW scheme dependence 
still remains when the complete one-loop corrections are 
supplemented by the leading higher-order corrections. To reduce 
the corresponding uncertainty, we need complete two-loop 
EW radiative corrections for the process under consideration.

The numerical results presented here 
were obtained using the Monte Carlo
generator {\tt ReneSANCe} \cite{Sadykov:2023azk} and the
{\tt MCSANCee} integrator which allow one to evaluate of
arbitrary differential cross sections and to separate particular contributions.

\section{Funding}
\label{sec:funding}

The research was supported by the Russian Science Foundation, project No. 22-12-00021.
\label{sec:acknowledgements}


\providecommand{\href}[2]{#2}
\end{document}